\documentclass[aps,english,10pt,a4,superscriptaddress,onecolumn,nofootinbib]{revtex4}  
\usepackage{amsmath}
\usepackage{amsthm}
\usepackage{amssymb}
\usepackage{amsfonts}
\usepackage{bbm}
\usepackage{epsfig}
\usepackage{times}
\usepackage{babel}
\usepackage{color}
\usepackage{hyperref}
\usepackage{framed}
\usepackage{changes}

\begin{document}
\title{Quantum theory as a principle theory: insights from an information-theoretic reconstruction}
\author{Adam Koberinski}
\affiliation{Department of Philosophy and Rotman Institute of Philosophy, University of Western Ontario, London, ON N6A 5BY, Canada}
\author{Markus P. M\"uller}
\affiliation{Institute for Quantum Optics and Quantum Information, Austrian Academy of Sciences, Boltzmanngasse 3, A-1090 Vienna, Austria}
\affiliation{Department of Applied Mathematics, University of Western Ontario, London, ON N6A 5BY, Canada}
\affiliation{Department of Philosophy and Rotman Institute of Philosophy, University of Western Ontario, London, ON N6A 5BY, Canada}
\affiliation{Perimeter Institute for Theoretical Physics, Waterloo, ON N2L 2Y5, Canada}

\begin{abstract}
We give a condensed and accessible summary of a recent derivation of quantum theory from information-theoretic principles, and use it to study the consequences of this and other reconstructions for our conceptual understanding of the quantum world. Since these principles are to a large extent expressed in computational terminology, we argue that the hypothesis of ``physics as computation'', if suitably interpreted, attains surprising explanatory power. Similarly as Jeffrey Bub and others, we conclude that quantum theory should be understood as a ``principle theory of information'', and we regard this view as a partial interpretation of quantum theory. We outline three options for completion into a full-fledged interpretation of quantum theory, but argue that, despite their interpretational agnosticism, the principled reconstructions pose a challenge for existing $\psi$-ontic interpretations. We also argue that continuous reversible time evolution can be understood as a characteristic property of quantum theory, offering a possible answer to Chris Fuchs' search for a ``glimpse of quantum reality''.
\end{abstract}

\date{July 12, 2017}

\maketitle

\section{Introduction}
Quantum theory has a long history of axiomatization. Von Neumann \cite{vonNeumann} began by placing Schr\"{o}dinger's wave mechanics and Heisenberg's matrix mechanics within a more general Hilbert space framework. Hilbert space quantum theory has since become the dominant formalism, due in large part to its abstract and general characterization. Von Neumann's ``axioms''---and later modifications---focus on the representation of quantum systems within that formalism. As such, the axioms tend to serve as a recipe for constructing a mathematical model of a given quantum situation. These recipes leave much open in terms of physical interpretation and explanation. For example, a quantum state is represented by a ray in a Hilbert space, and undergoes unitary time evolution within the space; there is no explanation regarding \textit{why} quantum systems evolve the way they do, or \textit{what} a quantum state is.

This explanatory gap represents one of the main motivations for the proliferation of different \textit{interpretations} of quantum theory, including the historically dominant Copenhagen interpretation~\cite{SEPCopenhagen}, Bohm's pilot wave theory~\cite{DurrBohm}, Everettian many worlds interpretations~\cite{ManyWorlds}, dynamical collapse theories~\cite{SEPDynCol} and several others.\footnote{The citations listed above are not to the originators of the interpretations; rather, they provide general overviews of each interpretation as each currently stands.} A substantial part of this research is characterized by a common goal: given the accepted mathematical structure of quantum theory, provide a comprehensible ontology with only minimal modification to its empirically confirmed predictions. A hoped-for consequence of this project is that the interpretation will solve certain conceptual problems, like the measurement problem. A major difference among interpretations is the role that the wave function plays in their ontology. Following~\cite{LeiferTalk,Leifer,Cabello}, we can distinguish ``$\psi$-ontic'' interpretations, which regard the quantum state as an element of reality, from ``$\psi$-epistemic'' interpretations which see the quantum state as a ``representation of knowledge, information, or belief''~\cite{Leifer}. Except for the Copenhagen interpretation, all of the traditional interpretations listed above fall into the $\psi$-ontic camp.\footnote{Feintzeig~\cite{Feintzeig} has questioned the applicability of the ontological models framework to Bohmian mechanics. For the purpose of our argumentation, however, it is sufficient to consider a qualitative distinction between $\psi$-epistemic and $\psi$-ontic theories, since none of our arguments rely on the formal mathematical definition of these notions in terms of the ontological models framework.}

The appearance of further $\psi$-epistemic interpretations (including the views of Wheeler~\cite{Wheeler}, Brukner and Zeilinger~\cite{Brukner,BZ2009,Zeilinger}, and QBism~\cite{QBism}) went hand in hand with a broader and more pragmatic development in the foundations of quantum mechanics. Physicists such as David Deutsch~\cite{Deutsch,Deutsch89} and Richard Feynman~\cite{Feynman} became aware of the potential computing power possessed by quantum systems. They realized that entanglement was not just a peculiarly quantum phenomenon, but had interesting uses as a computational resource. Over the late 1980s and throughout the 1990s the field of quantum computation saw major progress, guided by the idea of transmitting and harnessing \textit{quantum information} rather than classical information. The promise of improved computing power with quantum computers was a driving force behind the development of quantum information theory, which in turn shed new light on the field, leading to greater insight into the structure of quantum theory.

A key figure in merging quantum information theory with foundational work in quantum mechanics was John A.\ Wheeler~\cite{Wheeler}, whose ``It from Bit'' doctrine---that the fundamental \textit{stuff} of the universe is information---led to attempts to provide a coherent ontological story for quantum theory in information-theoretic terms. Clifton, Bub, and Halvorson (CBH)~\cite{CBH} appeared to have gone a long way towards this goal in their 2003 paper. They assume a background theory space characterized by $C^*$-algebras, which are general enough to include both classical and quantum theories, and derive the algebraic structure of standard quantum theory from information-theoretic postulates. In response to Wheeler, CBH \emph{``are suggesting that quantum theory be viewed\ldots as a theory about the possibilities and impossibilities of information transfer''} (p.\ 1563).

One of the major drawbacks of the CBH approach (later acknowledged by one of the authors~\cite{BananaWorld}) is that it relies on a $C^*$-algebraic framework for theories.\footnote{An early criticism of CBH's $C^*$-algebraic framework is due to Duwell~\cite{Duwell}. While we broadly agree with most of his conclusions, we also disagree with some of his arguments. For example, we think that it is not necessary for the information-theoretic constraints to be generally true; for the explanatory power of a reconstruction, it is sufficient that they are true in the regime of applicability of standard quantum theory (excluding any conjectured hidden-variables regime). The same reasoning applies to expectation value additivity.} Though the framework is general enough to include both classical and quantum theories,\footnote{This includes more general quantum physics, such as quantum field theory and quantum statistical mechanics. The former is somewhat controversial, given that only idealized, nonphysical models of quantum field theory have been constructed in the $C^*$-algebra framework. The majority view in the field is that the framework \textit{can} accommodate more realistic models, though their construction is elusive. Fraser and Wallace \cite{Fraser09,Wallace,Fraser11} have debated the relationship between algebraic and conventional approaches to quantum field theory.} its structure is rather restrictive, and builds in many assumptions that are crucial to quantum theories. Given the CBH reconstruction of quantum theory, the natural question to ask is, ``Why $C^*$-algebras for physical theories, rather than something else?''

Independently from the philosophers' attempts, physicists and mathematicians had already developed an operational framework that generalizes quantum theory in a much more universal way than the CBH approach. This framework dates back to at least the 1960s, and has been known in the mathematics and physics communities under several different names (order unit/base norm spaces, convex-operational framework, or \emph{generalized probabilistic theories} (GPTs), the name that we will use in this paper), and in slightly different versions. The goal of this framework is to capture all conceivable situations in physics that can be cast in operational language, in particular typical laboratory situations where there is a \emph{preparation procedure}, followed by a \emph{measurement} (and possibly \emph{transformation procedures} in between both). Ideally, the framework is so general that there are no constraints on its applicability,\footnote{It is a misunderstanding to say that this framework would not apply in the relativistic regime, or to say that the reconstructions of quantum theory in the GPT framework (as described in Section~\ref{SecDerivingQuantumTheory} for example) would be inherently ``non-relativistic''. Instead, the GPT framework itself is ignorant about the structure of spacetime: it can apply both to relativistic or non-relativistic scenarios, or to even more general situations (exactly as the Hilbert space and $C^*$-algebraic frameworks, which are special cases). Applying a certain GPT in a specific relativistic situation will in general involve \emph{additional effort}, for example by modelling Lorentz transformations in the corresponding theory, but there is no apriori obstruction to do so. The only practical restriction in most current work on GPTs is the limitation to finite-dimensional state spaces for technical reasons, which may have to be overcome for some applications.} other than the stipulation that it ought to describe the \emph{statistical or probabilistic} properties of a physical theory, and nothing else.

We will describe this framework\footnote{Philosophically inclined readers with an interest in some more details may find the exposition by Myrvold~\cite{Myrvold} helpful, which includes a comparison of the GPT and $C^*$-algebraic approaches.} in more detail in Section~\ref{secGPT}; for now, it is important to note that the GPT framework is comprehensive enough to allow for a variety of conceivable physical phenomena that are more general than those predicted by quantum theory. For example, some of the theories in this framework admit stronger forms of nonlocality than allowed by any version of quantum theory~\cite{PopescuRohrlich,Barrett2007}, a notion of ``higher-order interference''~\cite{Sorkin} that has already been tested experimentally~\cite{Sinha}, and beyond-quantum computational and information-theoretic properties like trivial communication complexity~\cite{vanDam} or a violation of information causality~\cite{InfCaus}. This framework provides a much more general starting point than the CBH reconstruction, from which to impose physically motivated postulates to single out quantum theory. In particular, instead of asking \emph{``Why quantum mechanics, rather than classical mechanics?''}, the focus has shifted towards asking a different question: \emph{``Why quantum mechanics, and not something possibly even more general?''}

Broadly speaking, there have been two different approaches to deriving quantum theory within the GPT framework. The first approach, pioneered by Lucien Hardy~\cite{Hardy2001},\footnote{However, there has been a long tradition of reconstruction attempts before Hardy's work; for a historical overview, see e.g.~\cite[Sec.\ 1.2]{Hardy2011} and the references therein.} starts with the (large) set of all probabilistic theories with arbitrary sets of states, measurements, and transformations (i.e.\ dynamics), and then imposes a small set of physically reasonable postulates on top of that framework, followed by a proof that quantum theory is the unique theory in the framework that satisfies these postulates. This approach has led to several fully successful reconstructions of quantum theory (e.g.~\cite{Hardy2001,DakicBrukner,Masanes,Chiribella,Hardy2011,PNAS,Hoehn,HoehnWever}); one major goal of this paper is to describe one of these reconstructions~\cite{PNAS} in an accessible way, and to discuss its implications (see Section~\ref{SecDerivingQuantumTheory}). Another (sometimes called ``device-independent'') approach disregards dynamics, and tries to single out the set of (static) \emph{quantum correlation tables} via some set of principles from the larger set of all non-signalling correlations, cf.\ Subsection~\ref{SubsecTwoGbits}. Despite important successes, this approach has not been able to \textit{exactly} single out quantum correlations so far~\cite{AlmostQuantum}. In Section~\ref{SecDerivingQuantumTheory}, we will speculate that this may tell us something important about physics.

Following the former approach, one has a framework  that admits the \textit{full} encoding of the essence of quantum theory into a set of simple postulates, singling out quantum theory from the ``landscape'' of all probabilistic theories. What is remarkable is that these postulates are to a large extent expressed in \emph{computational} terminology. As explained in detail in Section~\ref{SecDerivingQuantumTheory}, quantum theory is singled out by postulating that the state and time evolution of every physical system can be reversibly encoded into a number of interacting ``universal bits''; time evolution can be reversible and continuous; global states are uniquely determined by their local properties and their correlations; and one universal bit can carry one binary unit of information and not more.

These are properties which are directly linked to the possibility of having a \emph{universal computing machine}, like the quantum Turing machine, which is constructed in a modular way by composing a large number of universal bits. This universal machine is in principle able to simulate the time evolution of any physical system whatsoever, thereby rendering the empirically accessible content of any physical system---that is, its (quantum) state---ultimately ``substrate-independent''. It is interesting to see that this notion of ``universal computation'', as formalized in our postulates, is powerful enough to uniquely determine the state space, time evolution, and possible measurements (and thus also other properties like the maximal amount of non-locality) of quantum theory. In this sense, the hypothesis of ``physics as computation'', interpreted in a suitable way, demonstrates remarkable explanatory power.

The remainder of this paper is organized as follows. Section~\ref{SecThesis} outlines our view that information-theoretic reconstructions of quantum theory provide a fruitful, albeit only partial interpretation of quantum theory. We describe several alternatives for arriving at a full-fledged interpretation of quantum theory, using the information-theoretic reconstructions as a starting point. Furthermore, we describe why we think that the reconstructions pose a challenge to existing $\psi$-ontic interpretations. In Section~\ref{secGPT} we introduce the GPT framework, and highlight some of the generalizations beyond quantum theory by examining the gbit and its possible correlations. This is, in some sense, the most general fundamental unit of information within the GPT framework. In Section~\ref{SecDerivingQuantumTheory} we present the postulates that allow us to single out quantum theory uniquely from the space of GPTs, as first shown by one of the authors in~\cite{PNAS}. In doing so, we highlight the physical significance of the postulates, and some of their potential conceptual consequences. We conclude that information theory---specifically, constraints arising from universal computation---teaches us a lot about the structure of quantum theory. This goes a long way toward answering the question, ``Why the quantum?''

\section{Partially interpreting quantum theory as a principle theory}
\label{SecThesis}
We begin with a distinction first made by Einstein between \textit{principle} and \textit{constructive} theories. Reflecting on his methods for developing the special and general theories of relativity (SR and GR, respectively), Einstein contrasted his principle approach with the usual constructive methods in physics:
\begin{quote}
We can distinguish various kinds of theories in physics. Most of them are constructive. They attempt to build up a picture of the more complex phenomena out of the material of a relatively simple formal scheme from which they start out. Thus the kinetic theory of gases seeks to reduce mechanical, thermal, and diffusional processes to movements of molecules---i.e., to build them up out of the hypothesis of molecular motion. When we say that we have succeeded in understanding a group of natural processes, we invariably mean that a constructive theory has been found which covers the processes in question. Along with this most important class of theories there exists a second, which I will call `principle theories.' These employ the analytic, not the synthetic, method. The elements which form their basis and starting-point are not hypothetically constructed but empirically discovered ones, general characteristics of natural processes, principles that give rise to mathematically formulated criteria which the separate processes or the theoretical representations of them have to satisfy. Thus the science of thermodynamics seeks by analytical means to deduce necessary conditions, which separate events have to satisfy, from the universally experienced fact that perpetual motion is impossible~\cite[p.\ 228]{EinsteinLetter}.
\end{quote}

Others have noticed the similarity between information-theoretic reconstructions of quantum theory and SR \cite{HarriganSpekkens,Zeilinger,Felline,BrownTimpson}, and Bub has been vocal about the utility of information theory for making quantum mechanics a principle theory \cite{Bub2005,CBH,BananaWorld}. Unlike constructive theories, principle theories explain by showing that features of the world are deductive consequences of a small set of general principles. This leads us to formulate the following thesis, which we will elaborate in more detail in the remainder of this section:\\

\textbf{Thesis:} \textit{Information-theoretic reconstructions provide a partial interpretation of quantum theory as a \emph{principle theory of information},\footnote{This is very closely along the lines of Jeffrey Bub's~\cite{Bub2005,CBH} suggestion.} by identifying a small set of information-theoretic principles that render our world a quantum one. This leaves several alternatives for extending to a full-fledged interpretation, some of which we describe below. Furthermore, the reconstructions represent a challenge for existing ``$\psi$-ontic'' interpretations of quantum theory, by highlighting a relative deficiency of those interpretations in terms of their explanatory power.}\\

Standard textbook formulations often phrase quantum theory as a set of axioms (e.g.\ that states are given by rays on a Hilbert space, the Born rule, etc.), but these axioms are purely mathematical and abstract, and thus do not correspond to the kind of physical principles that one would expect from a principle theory in Einstein's sense. Furthermore, the standard formulation also does not give an account in the sense of a constructive theory, and as such is ``doubly unsatisfactory''. On the other hand, if the quantum formalism is derived from simple physical postulates, and cast as a principle theory based on information, then one gains an explanation in the sense that quantum theory is found to be a unique consequence of easily understandable constraints. The mathematical ingredients therefore become less arbitrary, and provide a partial interpretation in terms of the role they play in satisfying the more comprehensible principles.

Nonetheless, the information-theoretic reconstructions do \emph{not} provide explanatory power in an even stronger sense: they do not typically tell us \emph{what quantum states are}, or \emph{what is really going on in the world} when we perform a Bell experiment, for example. In particular, information theory provides a foundation for understanding quantum theory as a principle theory, but \textit{not} as a constructive theory. However, we highlight three alternatives for a potential generalization to a full interpretation of quantum theory, using information-theoretic reconstructions as a starting point.\footnote{Laura Felline~\cite{Felline} has given a detailed analysis of the notion of explanation that applies to information-theoretic reconstructions of quantum theory, and her conclusions are to a large extent compatible with what we outline in this paper. Felline has also described two of our options (for a full-fledged interpretation), namely the possibility of a `structural' explanation (in conjunction with the rejection of the necessity of causal explanations) and the option to find a constructive successor of quantum theory.}

First, quantum information theory coupled with ontic structural realism (OSR)~\cite{Ladyman} could provide a full-fledged interpretation of quantum theory, by rejecting the ultimate need for an entity-based constructive account of a fundamental theory of physics. This would consist in an ontology of structural relations in some sense---simply of the relational structure uniquely picked out by the information-theoretic postulates from the space of GPTs. This option denies the need for a constructive account of quantum theory, though it does not rule out the possibility of discovering a constructive successor to quantum theory, in particular since ontological stability across theory change is a characteristic of OSR.\footnote{Lucas Dunlap's current work is concerned with a detailed analysis of this approach; see e.g.\ his talk ``The Information-Theoretic Interpretation of Quantum Mechanics and Ontic Structural Realism'' at \emph{Foundations 2016: The 18th UK and European Conference on Foundations of Physics}, London School of Economics. Note that quantum theory has been an important motivation in the development of OSR~\cite{LadymanSR}, but it has also been argued that OSR in itself is not sufficient to constitute an interpretation of quantum mechanics~\cite{Esfeld}.}

Second, one could adopt a subjective view of quantum theory, what Cabello calls participatory realist interpretations of quantum theory~\cite{Cabello}. Though these fall into various forms, subjective interpretations of quantum theory all share the common point of view that quantum theory is not directly about properties of our world. Most subjective interpretations (including the Copenhagen interpretation,\footnote{At least Heisenberg's view can be characterized in this way~\cite{ObjQM}. ``\textit{The} Copenhagen interpretation'' is a bit misleading, as there is only a loose set of common elements amongst thinkers like Bohr and Heisenberg. See~\cite{SEPCopenhagen} for a more nuanced discussion.} the view of Brukner and Zeilinger~\cite{Zeilinger,BZ2009,BZ2003}, and Wheeler's interpretation~\cite{Wheeler}) claim that quantum theory is a theory about our \textit{knowledge} of the world but not the world itself. Another possibility is given by the ``Quantum Bayesian'' (QBist) interpretation that quantum theory is an extension of Bayesian rules for subjective, rational \textit{belief} updating~\cite{QBism,QBism2}. Like the first option, this route ``explains away'' the need for a constructive account of quantum theory, although some of its proponents (in particular, QBists) express the hope for insights into the actual objective world as a major motivation for their approach.

The final option is inspired by the quote above highlighting Einstein's thoughts on principle and constructive theories. Some might agree with Einstein's view that successful constructive theories carry more explanatory weight, and would prefer a constructive version of quantum theory. Consequently, the third option is to come up with a constructive successor of quantum theory, describing a regime of our world that is currently empirically inaccessible, but which gives rise to the information-theoretic principles (and thus indirectly to quantum theory) after some approximation or coarse-graining. A paradigmatic historical example, as explained by Einstein, is the kinetic theory of gases as a constructive successor of thermodynamics. It is encouraging to see how, at least in this instance, the development of a constructive theory was guided and constrained by the presence of a principle theory. Thus, one might hope that the information-theoretic reconstructions could serve as a similar guideline in the case of quantum theory. In fact, thermodynamics and kinetic theory ended up working \textit{together} to form a fuller, more explanatorily powerful framework for the properties of bulk materials. We consider this additional power---in terms of prediction, or unification---a crucial signature of success of the constructive theory, and a main reason to adopt it in addition to the bare principle theory.

This third option seems to be particularly attractive for proponents of ``non-subjective'' $\psi$-epistemic interpretations of quantum theory (or ``$\psi$-epistemic type I'' in Cabello's nomenclature, i.e.\ those that ``view the quantum state as representing knowledge of an underlying objective reality''~\cite{Cabello}), insofar as they regard their interpretations as an open research program. Proponents of this view would agree that a quantum state describes our (limited) \emph{information} about the world, and could thus be happy with a principled formulation of quantum theory in information-theoretic terms as a starting point. They could see the information-theoretic reconstructions as a constraining guideline for the development of the sought-for constructive underpinning of quantum theory, in particular since they would consider these information-theoretic principles as more directly related to properties of the real world than people in the subjective camp would.

We note that these options are neither completely exhaustive, nor mutually exclusive, but they all go beyond what is directly provided by any reconstruction. However, even if one takes \textit{only} the partial interpretation that a principle theory provides, we think that its explanatory power is sufficient to pose a challenge to proponents of more traditional $\psi$-ontic interpretations. None of Bohmian mechanics, Everettian quantum theory or collapse theories fill the explanatory role of a principle theory. For example, Everettian quantum theory does \emph{not} start with a broad general framework of ``theories of many worlds'', put simple principles on top of that, and prove that quantum theory is the unique theory of many worlds that satisfies these principles.

One can certainly object that $\psi$-ontic theories are not meant to be principle theories in the first place---their ambition is to exhaustively describe what is going on in a quantum world, and as such they would more accurately be described as candidate constructive theories (possibly with Bohmian mechanics coming closest to Einstein's vision of a constructive theory). Nevertheless, given that they do not provide the specific explanatory power of the information-theoretic derivations that we have just described, one should demand that they give us something else in replacement in order to be considered successful: some additional element of empirical or theoretical success that goes \emph{beyond} the standard formulation of quantum theory. Clearly, the aforementioned example of kinetic theory (underpinning thermodynamics) satisfies this demand in a quite spectacular way, but we do not think that any of the existing $\psi$-ontic interpretations of quantum theory currently does.

Now the question of \emph{what exactly} would or should constitute an empirical or theoretical success which would cause physicists to accept a given $\psi$-ontic theory is clearly a difficult and controversial question. Some supporters of $\psi$-ontic theories will see the \emph{realist} worldview that these theories provide as enough of a success to accept them. However, in contrast to kinetic theory, this acceptance currently comes at the price of \emph{not also} being provided with additional elements of explanation or prediction which, for example, could help to empirically validate the specific claims of one $\psi$-ontic theory (say, Bohmian mechanics) over another one (say, the many-worlds interpretation). What the information-theoretic derivations do is to emphasize, and extend, this explanatory deficiency: they demonstrate how a (partial) interpretation of quantum theory as a principle theory of information can provide novel explanations of the quantum formalism that are not supplied by the current $\psi$-ontic theories\footnote{Since the information-theoretic reconstructions are largely agnostic about the physical substrate underlying quantum theory, they are in principle entirely compatible with existing $\psi$-ontic interpretations. Our objection is that current $\psi$-ontic interpretations do not empirically supplement the principle theory in a similar way as the kinetic theory of gases empirically supplements thermodynamics.}.

Is this argument also a challenge for $\psi$-epistemic interpretations? We do not think so. Interpretations that treat the quantum state as a state of knowledge or belief are conceptually more closely related to the view of quantum theory as a principle theory of information, which has led some physicists (e.g.~Brukner\footnote{For example, at the conference ``Quantum Theory: from Problems to Advances'' in V\"axj\"o, Sweden (June 9-12, 2014), Brukner argued as follows: ``The \emph{very idea of quantum states as representatives of information} --- information that is sufficient for computing probabilities of outcomes following specific preparations --- \emph{has the power} to explain why the theory has the very mathematical structure it does. This in itself is the message of the reconstructions.''}) to argue that the success of the latter is evidence for the validity of the former. Another major difference lies in the fact that the $\psi$-ontic interpretations make much more specific claims about the real world. For example, Bohmian mechanics stipulates that the universe is in some (unknown) configuration, which (together with a pilot wave) evolves according to specific differential equations. On the other hand, while QBists view quantum states as subjective states of belief, they are rather agnostic about the underlying real world that gives rise to these beliefs, and see the question of ontology as much more of an open research program. Thus, different demands for predictive power in the empirical regime follow from a simple doctrine: that concrete, specific claims about the real world should ultimately be backed up by successful empirical predictions which do not also arise identically from rival theories.

\section{The GPT framework: General no-signalling theory (boxworld)}
\label{secGPT}
The empirical content of quantum theory lies in the statistical prediction of measurement outcomes: if the preparation of a quantum system in state $\rho$ (a density matrix) is followed by an $n$-outcome measurement with projection operators $P_1,\ldots,P_n$, then the probability of the $i$th outcome is given by ${\rm tr}(\rho P_i)$. This is the minimal content on which all interpretations of quantum mechanics agree. The framework of generalized probabilistic theories (GPTs) allows one to describe basically \emph{any} theory that fits within this basic operational prescription, namely that a preparation procedure is followed by a measurement (and possibly a transformation procedure before the measurement takes place), yielding one of several possible outcomes probabilistically. Classical probability theory\footnote{Here, ``classical probability theory'' is not meant to describe the scientific discipline, but concretely stands for the state space of discrete probability distributions, in contrast to quantum theory where the states are density matrices. That is, it is the technical notion of a specific GPT where the set of states $\Omega$ is the set of probability vectors $(p_1,\ldots,p_n)$ (where $n\in\mathbb{N}$ is fixed), $p_i\geq 0$, $\sum_i p_i=1$, and the reversible transformations are the permutations of entries. Geometrically, these (``classical'') state spaces are simplices.} and quantum theory are special cases of GPTs, but there are many others which are neither classical nor quantum.

Here, we will \emph{not} give a complete formal definition of the GPT framework; interested readers can find such a definition, for example, in~\cite{Barrett2007,Hardy2001}. Instead, we focus on a simple illustrative example of one specific non-classical and non-quantum theory within this framework, sometimes called ``general no-signalling theory''~\cite{Barrett2007}, or, more colloquially, ``boxworld''.

First we introduce a single system in boxworld, sometimes called the ``gbit'' (generalized bit), and we will see in what way it differs from quantum theory's ``qubit''. Then we describe a composite state space of two gbits, which turns out to be the infamous ``no-signalling polytope''. Finally, we discuss some of its properties, and point out crucial differences with the state space of two quantum bits. This example will also serve to illustrate the use and the meaning of the information-theoretic postulates which allow one to reconstruct quantum theory, as we will discuss in Section~\ref{SecDerivingQuantumTheory}.

\subsection{The ``gbit'': a single generalized bit in boxworld}
\label{SubsecGbit}
Consider a single spin-$1/2$ particle, for example an electron. Quantum physics allows us to prepare the spin degree of freedom of that particle in any quantum state (described by a $2\times 2$ density matrix $\rho$), and later on to measure the spin in any spatial direction $\vec n$ that we choose---for example, by setting up a Stern-Gerlach device with a magnetic field gradient in direction $\vec n$. The measurement outcome will always be either spin-up or spin-down. Since we always have two possible outcomes, this represents an elementary binary alternative within quantum theory---a ``qubit''.

Now imagine a world where there are particles that behave in a somewhat similar way: we can prepare them in some state $\omega$ (for now let us remain silent about what mathematical object this is), and later on decide to perform a two-outcome measurement (which we will denote by $x$). However, let us assume that there are only \emph{two} possible choices of two-outcome measurement, denoted $x=0$ and $x=1$. This is in contrast to the quantum particle, where we have infinitely many possible two-outcome measurements, corresponding to the different choices of quantization axis $x=\vec n$. Let us denote the possible outcomes by $\uparrow$ and $\downarrow$; in the quantum case, these correspond to ``spin-up'' and ``spin-down'' respectively.

If we decide to perform measurement $x$ on one of these hypothetical particles, we get outcome $a$ with some probability $p(a|x)$. Clearly $p(\uparrow|x)+p(\downarrow|x)=1$ both for $x=0$ and $x=1$. This means that the two real numbers $p(\uparrow|0)$ and $p(\uparrow|1)$ tell us everything that we need to know to predict the outcome of any measurement that we could possibly perform on a particle. We can thus write down the state $\omega$ simply as
\[
   \omega=\left(\strut p(\uparrow|0),p(\uparrow|1)\right).
\]
Now we make one further assumption: namely, that the hypothetical physics that describes our hypothetical particle does not restrict these two probabilities in any way. In other words, we imagine that for \emph{any} given choice of these two numbers (in the unit interval $[0,1]$), we can in principle prepare a particle in the corresponding state $\omega$. Then the set of all possible states, the \emph{state space}, can be visualized as all the points in two dimensions with coordinates between $0$ and $1$, that is, the unit square. Note that the state itself does not tell us anything about \emph{what that particle is}, or \emph{what happens during a measurement}. All it does is to allow us to compute the outcome probabilities of all possible future measurements---it formalizes the operational content of the physics of the hypothetical particle, and not more.

\begin{figure}
\begin{center}
   \includegraphics[width=0.4\textwidth]{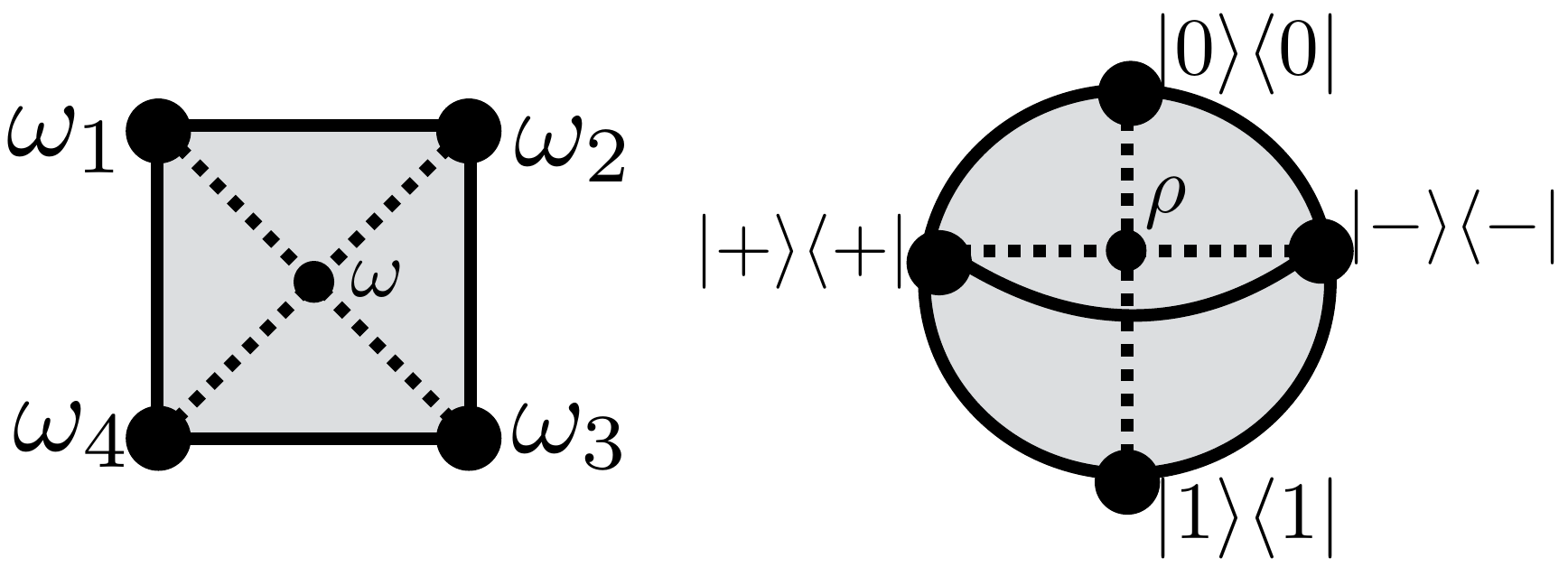}
	\caption{The state spaces of a gbit (left) and a qubit (right). The gbit state space has four pure states (that is, extremal points of the convex state space), namely $\omega_1=(0,1)$, $\omega_2=(1,1)$, $\omega_3=(1,0)$ and $\omega_4=(0,0)$. For any given state $\omega$ (for example $\omega=(\frac 1 2, \frac 1 2)$), the decomposition into pure states is not unique (for this example, $\omega=\frac 1 2 \omega_1+\frac 1 2 \omega_3 = \frac 1 2 \omega_2+\frac 1 2 \omega_4$). This is a signature of nonclassicality, and is also present in the quantum bit. For example, for the maximally mixed state $\rho=\frac 1 2 \mathbf{1}$, we have $\rho=\frac 1 2 |0\rangle\langle 0|+\frac 1 2 |1\rangle\langle 1|=\frac 1 2 |+\rangle\langle+|+\frac 1 2 |-\rangle\langle -|$, where $|\pm\rangle=\frac 1 {\sqrt{2}}(|0\rangle \pm |1\rangle)$. The qubit state space has infinitely many pure states (all states on its boundary).}
	\label{fig_gbitqubit}
\end{center}
\end{figure}

Figure~\ref{fig_gbitqubit} illustrates this, and compares it to the quantum bit. For a qubit, the state can be described by a density matrix
\[
   \rho=\frac 1 2 \left(
      \begin{array}{cc}
         1+a_1 & a_2-i a_3 \\
         a_2+i a_3 & 1-a_1	
      \end{array}
   \right).
\]
Density matrices must have non-negative eigenvalues. Since the two eigenvalues are $\frac 1 2 \left(1\pm \sqrt{a_1^2+a_2^2+a_3^2}\right)$, we obtain the condition $|\vec a|\leq 1$ for the vector $\vec a:=(a_1,a_2,a_3)$. We can thus visualize the quantum bit as the unit ball in three dimensions, which is known as the \emph{Bloch ball}. As we see in Figure~\ref{fig_gbitqubit}, the gbit and the qubit state spaces thus look very different, but both have the important property of \emph{convexity}. This formalizes the idea of a ``mixed state''. Imagine a procedure where one tosses a fair coin, and, depending on the outcome, prepares either state $\omega$ or another state $\varphi$. Repetition of this procedure will statistically correspond to the preparation of the state $\frac 1 2 \omega+\frac 1 2\varphi$---a \emph{convex combination} of $\omega$ and $\varphi$. Therefore, convex combinations of states are also states, which forces state spaces to be convex sets. In principle, \emph{any} convex, closed, bounded set in a vector space over the real numbers\footnote{In the examples above, this vector space will be $\mathbb{R}^2$ (for the gbit) and the real vector space of Hermitian complex $2\times 2$ matrices (for the qubit). See~\cite[Chapter 1]{Holevo} for a detailed explanation why one can always without loss of generality represent state spaces as subsets of a real vector space.} corresponds to a valid state space in the framework of GPTs.

Many important physical properties of a conceivable physical system can be inferred from the shape of its state space. First, one can see that both the gbit and the qubit are \emph{nonclassical} in the sense that states have in general more than one convex decomposition into pure states. Classically, a state corresponds to a probability distribution over mutually distinguishable alternatives (such as probability $p$ for ``heads'' and $1-p$ for ``tails'' for a coin), and there is always a unique way to decompose that state into its alternatives. However, this is not true for the qubit nor for the gbit, as Figure~\ref{fig_gbitqubit} illustrates.

Another important property of a state space is to ask what kinds of \emph{reversible transformations} are possible. We can think of a ``transformation'' as any operation that we apply to the particle, e.g.\ switching on a magnetic field for some time, or performing a ``gate'' in a computation. In particular, \emph{time evolution} from some initial state $\omega_i$ to some final state $\omega_f$ is an example of a transformation $T$, i.e.\ $\omega_f=T(\omega_i)$. To preserve the statistical interpretation of mixtures, transformations $T$ must necessarily be affine-linear~\cite{Holevo,Barrett2007}. A transformation is \emph{reversible} if it can be inverted, and if its inverse $T^{-1}$ is a transformation too. In the quantum case, the reversible transformations are exactly the \emph{unitary transformations} $U$, mapping an initial state $\rho_i$ to a final state $\rho_f=U\rho_i U^\dagger$. In the Bloch ball picture, unitary transformations correspond to \emph{rotations} of the Bloch ball. In other words, for every unitary $U$ there is a $3\times 3$ rotation matrix $R$ such that the final Bloch vector $\vec a_f$ is related to the initial Bloch vector $\vec a_i$ via $\vec a_f=R\vec a_i$. This illustrates the general fact that reversible transformations in a GPT are \emph{affine-linear symmetries} of the state space.

Clearly, a three-dimensional ball can be continuously rotated, which means that we can have \emph{continuous reversible time evolution} (for example, according to the Schr\"odinger equation) of the form $\vec a(t)=R(t)\vec a_i$, where $t\in\mathbb{R}$ denotes time. On the other hand, the only possible reversible transformations on a gbit are the symmetries of the square---rotations by multiples of $90^\circ$ and reflections. Time evolution of the state would have to pass in discrete steps; continuous reversible evolution is impossible. This is related to the fact that the qubit is ``round'' and the gbit is not; or, more formally, that the qubit has \emph{infinitely many pure states}, while the gbit has only four pure states (the corners).

\subsection{Two gbits: the no-signalling polytope}
\label{SubsecTwoGbits}
If we have two distinguishable quantum systems, described by Hilbert spaces $\mathcal{H}_A$ and $\mathcal{H}_B$, then there is always a unique corresponding composite quantum system, namely the one that is described by the tensor product Hilbert space $\mathcal{H}_A\otimes \mathcal{H}_B$. General state spaces in the GPT framework can be composed analogously, with the crucial difference that \emph{there are in general infinitely many possibilities on what the composite state space can be}. In more detail, if $\Omega_A$ and $\Omega_B$ are two state spaces (for example, gbits), then any state space $\Omega_{AB}$ is considered as a valid composite, as long as it satisfies the following postulates:
\begin{itemize}
	\item \textbf{Local preparations:} For every two states $\omega_A\in\Omega_A$ and $\omega_B\in\Omega_B$, there is a state in $\Omega_{AB}$ that corresponds to the independent local preparation of $\omega_A$ and $\omega_B$. Here we denote this state by $\omega_A\omega_B$.
	\item \textbf{Local measurements:} For every measurement on $\Omega_A$ that has possible outcomes $a_i$, and measurement on $\Omega_B$ that has possible outcomes $b_j$, there is a joint measurement on $\Omega_{AB}$ with pairs of outcomes $(a_i,b_j)$ that describes a situation where the two measurements are performed independently on $\Omega_A$ and $\Omega_B$. Denoting the probability of an outcome $a_i$ if one measures on state $\omega_A$ by $p_{\omega_A}(a_i)$, the corresponding measurement satisfies $p_{\omega_A\omega_B}(a_i,b_j)=p_{\omega_A}(a_i)p_{\omega_B}(b_j)$.
	\item \textbf{No-signalling:} If local measurements are performed on an arbitrary state in $\Omega_{AB}$, then the outcome probabilities on $\Omega_A$ do not depend on the choice of measurement on $\Omega_B$, and vice versa.
\end{itemize}
While these postulates are always taken as background assumptions in any GPT, the following postulate is sometimes introduced in addition:
\begin{itemize}
	\item \textbf{Tomographic locality:} States on $\Omega_{AB}$ are uniquely determined by the statistics of local measurements.
\end{itemize}
All these postulates are satisfied in quantum theory: if $\rho_A,\rho_B$ are density matrices on Hilbert spaces $\mathcal{H}_A$ and $\mathcal{H}_B$, then the corresponding local preparation is given by the density matrix $\rho_A\otimes\rho_B$ on $\mathcal{H}_A\otimes\mathcal{H}_B$. Similarly, if we have projectors\footnote{We can more generally consider \emph{positive operator-valued measures} where the $P_i$ are not necessarily projectors, but arbitrary positive-semidefinite operators.} $P_i^{(A)}$ and $P_j^{(B)}$ that define measurements on $\mathcal{H}_A$ and $\mathcal{H}_B$, then the projectors $P_i^{(A)}\otimes P_j^{(B)}$ define the corresponding local measurement on the composite quantum system. Since projectors of this product form span the full linear space of Hermitian matrices, their outcome probabilities ${\rm tr}(P_i^{(A)}\otimes P_j^{(B)} \rho_{AB})$ determine the state $\rho_{AB}$ uniquely, which implies tomographic locality. These postulates are also satisfied by classical probability theory, i.e.\ the state spaces of classical discrete probability distributions.

Now consider two gbit state spaces $\Omega_A$ and $\Omega_B$ (i.e.\ the same square state space twice, but with different labels). We define a composite state space $\Omega_{AB}$ in the following way:

\begin{center}
\textit{The composite state space $\Omega_{AB}$ shall contain all non-signalling probability tables of the form $p(a,b|x,y)$, where $x,y\in\{0,1\}$ denote the local choices of measurements, and $a,b\in\{\uparrow,\downarrow\}$ denote the two local outcomes.}
\end{center}
Intuitively, no-signalling means that the choice of experiment for one system has no effect on the statistics obtained for the second system. Mathematically, this is equivalent to the marginals $p(a|x)$ and $p(b|y)$ being well-defined and independent of the choice of experiment $y$ or $x$, respectively. This leads to
\begin{align}
p(a|x) &= \displaystyle\sum_{b} p(a,b|x,y) = \displaystyle\sum_{b} p(a,b|x,y') \; \mbox{ for all } a,x,y,y', \\
p(b|y) &= \displaystyle\sum_{a} p(a,b|x,y) = \displaystyle\sum_{a} p(a,b|x',y) \; \mbox{ for all }b,x,x',y.
\end{align}
These conditions are supplemented by the conditions of non-negativity, $p(a,b|x,y)\geq 0$, and normalization: $\sum_{a,b}p(a,b|x,y)=1$ for all $x,y$.

A state $\omega_{AB}$ of two gbits corresponds to a list of sixteen probabilities $p(a,b|x,y)$. However, the conditions above reduce the number of affinely independent probabilities to eight. This is similar to the single-gbit case, where we have started with four probabilities $p(a|x)$, but only two of them were affinely independent. Since we have a set of linear equalities and inequalities, the vectors (probability tables) that satisfy them---that is, the state space $\Omega_{AB}$---will form a polytope.

The vertices of this polytope fall into two types. The first type are local deterministic states, which correspond to local preparations where all probabilities are either zero or one:
\[
   p(a,b|x,y)=p(a|x)p(b|y), \mbox{ and }p(a|x),p(b|y)\in\{0,1\}.
\]
Since there are four possible choices for deterministic $p(a|x)$ (the four pure states of the gbit), and similarly for $p(b|y)$, we obtain sixteen vertices of this kind.

The second type of vertex is sometimes called a ``Popescu-Rohrlich box''~\cite{PopescuRohrlich}. There are eight vertices of this type. Let us only describe one of them; the other seven can be obtained from it by relabelling the measurement choices and outcomes. A PR-box state is
\[
   p(a,b|x,y)=\left\{
      \begin{array}{cl}
      	  1/2 & \mbox{if }(x,y)=(0,0)\mbox{ and } (a,b)\in\{(\uparrow,\uparrow),(\downarrow,\downarrow)\}\\
      	  1/2 & \mbox{if }(x,y)\in\{(0,1),(1,0),(1,1)\}\mbox{ and } (a,b)\in\{(\uparrow,\downarrow),(\downarrow,\uparrow)\}\\
      	  0 & \mbox{otherwise}.
      \end{array}
   \right.
\]
That is, if both parties decide to perform the $0$-measurement, then the outcomes are correlated; in all other cases, they are anticorrelated. The PR-box exhibits correlations that are impossible to obtain within quantum theory: a PR-box would violate the CHSH Bell inequality by more than any quantum state~\cite{PopescuRohrlich}. This illustrates one way in which general theories in the GPT framework can describe physics that differs from quantum physics.

\begin{figure}
\begin{center}
   \includegraphics[width=0.1\textwidth]{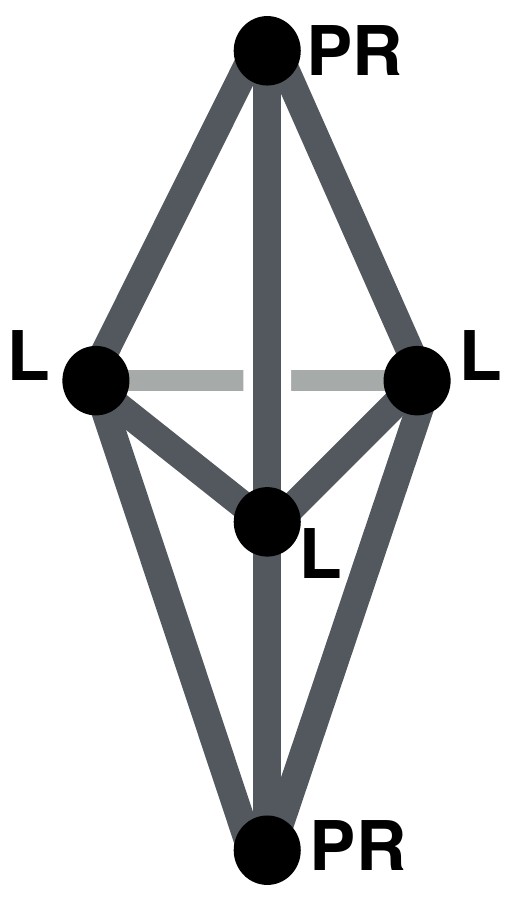}
	\caption{A caricature of the no-signalling polytope $\Omega_{AB}$ (which, in contrast to this schematic figure, is eight-dimensional and has $24$ vertices). There are two classes of pure states: local states (labelled ``L''), and states of PR-box type. Since no linear symmetry of this polytope can map a local vertex to a PR-box vertex, there cannot be any reversible transformations in boxworld that generate a PR-box state from a local state.}
	\label{fig_nspolytope}
\end{center}
\end{figure}

The state space $\Omega_{AB}$ is the polytope that is spanned by these twenty-four pure states (it is their convex hull, cf.\ Figure~\ref{fig_nspolytope}). It is eight-dimensional, and it has been extensively studied in quantum information theory~\cite{Barrett2007}. In our context, it will be particularly illuminating to study the possible \emph{reversible transformations} on $\Omega_{AB}$.

As we have argued above, every transformation of this kind must be an affine-linear symmetry of the state space. Thus, in order to find the reversible transformations, we need to mathematically classify these symmetries. This has first been done in~\cite{Gross}, where it was shown that the \emph{only} symmetries in boxworld (for any number of measurements, outcomes, and parties) correspond to relabellings of measurement outcomes on individual subsystems, and permutations of subsystems. In particular, \emph{there is no reversible transformation that maps a local deterministic state to a PR-box state}. Figure~\ref{fig_nspolytope} shows a caricature: if the state space were in fact identical to that in the figure, then it would be geometrically obvious that no linear symmetry can map one of the PR-box vertices to any of the local vertices. This is because the local and PR-box vertices as depicted in the caricature are geometrically inequivalent: every local vertex connects, via edges, to four other vertices, but every PR-box vertex is connected to three other vertices only.

The actual no-signalling polytope $\Omega_{AB}$ is more complicated than the caricature, but one can analyze its structure by using the PolyMake software~\cite{Polymake}. We can construct a description of $\Omega_{AB}$ within PolyMake, and then use it to count the number of vertices that are connected to a given vertex via edges. What we find is that the local vertices connect to 17 other vertices: 13 other local vertices and 4 PR-box vertices. The PR-box vertices, in contrast, connect to 8 other vertices, all of which are local vertices. In summary, this implies that \emph{reversible time evolution cannot map a deterministic state of two gbits to a PR-box state of two gbits, or vice versa}.

We can contrast this to the state space of two \emph{quantum} bits, $\Omega_{AB}^{QM}$. This is the state space of $4\times 4$ density matrices, which is not a ball (even though the qubit was), but a rather complicated $15$-dimensional convex set. In contrast to $\Omega_{AB}$, it has infinitely many pure states; some of them corresponding to local preparations (like $\rho_A\otimes\rho_B$ with $\rho_A,\rho_B$ pure states) and some of them not (like $|\psi_+\rangle\langle\psi_+|$, with $|\psi_+\rangle=\frac 1 {\sqrt{2}} (|\uparrow\downarrow\rangle+|\downarrow\uparrow\rangle)$ an entangled state). However, all these pure states are geometrically equivalent: if $\rho_{AB}$ and $\sigma_{AB}$ are any two pure states, then there is always a reversible transformation (namely, a unitary $U$) such that $\rho_{AB}=U\sigma_{AB}U^\dagger$.\footnote{For a more thorough introduction to the geometry of quantum state spaces, see Bengtsson and \.{Z}yczkowski~\cite{QMGeo}. For a more general overview of quantum information theory, see Nielsen and Chuang~\cite{QITtextbook}.} In particular, pure product states can be mapped to entangled states -- time evolution can create entanglement, and can do so continuously in time. This means that $\Omega_{AB}^{QM}$ is much more symmetric than $\Omega_{AB}$: every pure state looks like every other one.

Thus, the two-bit state spaces of quantum theory and boxworld have many features in common: they are non-classical (i.e.\ there is non-uniqueness of convex decompositions), they contain entangled states, and they satisfy the principle of tomographic locality. However, two-qubit quantum theory satisfies the following principle, while two-gbit boxworld does not:

\textbf{Reversibility:} Given any two pure states of the system, there is a reversible transformation that maps the first to the second.

\section{Deriving quantum theory}
\label{SecDerivingQuantumTheory}
Given the GPT framework, how do we arrive at quantum theory? One of us has shown (with further coauthors) in previous works that quantum theory is the unique consequence of a set of four principles, at least for finite-dimensional systems~\cite{PNAS,Masanes}. We outline the assumptions and principles behind the derivation (see the list below), and offer an analysis of the insight they provide into the structure of quantum theory. We start by discussing the principle of the \textit{Existence of an Information Unit}. Qualitatively, this means that there exists some universal information quantity (the ``universal bit''); given a sufficient number of these units, we can encode, model and decode the state of any physical system. 

More specifically, if we have any state $\omega$ of a physical system with finite-dimensional state space $\Omega$, and we would like to apply a transformation $T$ to arrive at a state $\varphi=T\omega$, then we can also achieve this by doing it in the following way:
\begin{itemize}
	\item reversibly encode $\omega\in\Omega$ into a state $\omega^{(n)}$ on the state space of $n$ universal bits (where $n$ is large enough),
	\item perform a reversible transformation $T^{(n)}$ on $n$ universal bits to arrive at $\varphi^{(n)}=T^{(n)}\omega^{(n)}$,
	\item reversibly decode the resulting state to obtain $\varphi\in\Omega$.
\end{itemize}
In quantum theory, we have such a universal bit, the \emph{qubit}. Furthermore, we can in principle simulate reversible evolutions (that is, unitary transformations) of \emph{any} physical system by encoding its state on a quantum computer, performing a unitary transformation on its quantum memory, and decoding the state back into the physical system in question. Postulate 3 below formulates this property in the GPT framework, without specifically assuming quantum theory. Consequently, \emph{classical probability theory}, where the states are discrete probability distributions, satisfies the principle too: the universal bit is simply the bit.

This can be interpreted as providing a necessary condition for nature being computable in a certain sense: whatever model of computation we may come up with (like, for example, the quantum Turing machine), it is necessarily a crucial aspect of that model to be ``modular'', i.e.\ to consist of a (large) collection of a small set of primitive building blocks. These are the universal bits. The scheme above says that \emph{all} physical systems and their dynamical evolutions can in principle be reliably simulated using a sufficient number of universal bits and their dynamical interaction, which admits simulation of physics on a computer. Note that this encoding-decoding-scheme, and its formulation in terms of the postulates below, does not require \emph{all} ingredients that are important for universal (quantum) computation; for example, it does not postulate that all dynamical evolutions can be approximated by a fixed finite ``universal gate set'' (which is the case in quantum theory, see Appendix 3 in~\cite{QITtextbook})---instead, this property follows as a consequence of the postulates. As a further consequence, since all systems can be encoded in a sufficient set of these information units, the task of characterizing the theory's state spaces is reduced to that of characterizing the state space of $n$ universal bits, for all $n\in\mathbb{N}$. In this sense, information and computability are central to deriving quantum theory.

The full set of principles are listed and described below. Together, they suffice to uniquely single out quantum theory:

\textit{\textbf{Theorem}~\cite{PNAS}: The unique GPT that satisfies the postulates below is quantum theory.}

\textit{That is, the state space of $n$ universal bits is exactly the state space of $n$ qubits: the states are the $2^n\times 2^n$ density matrices, the reversible transformations are the conjugations by unitary matrices, $\rho\mapsto U\rho U^\dagger$, and the possible measurements are the positive operator-valued measures (POVMs).
}
The postulates are:
\begin{itemize}
\item[1.] \textbf{Continuous Reversibility}. Given two pure states $\omega$ and $\varphi$, there is a family of reversible transformations $G(t)_{0\leq t\leq 1}$ such that $G(t)$ is continuous in $t$, $G(0)=\mathbf{1}$ and $G(1)\omega=\varphi$.
\end{itemize}
In other words, any pure state $\omega$ can be reversibly and continuously (``in time'', if $t$ is interpreted as a time parameter) transformed into any other pure state $\varphi$. This is a stronger version of the ``Reversibility'' postulate introduced in Subsection~\ref{SubsecTwoGbits}. It is well-motivated by experience with physical systems: states may be adjusted to any of a continuum of settings, and physical evolution is often continuous and reversible in time.
\begin{itemize}
\item[2.] \textbf{Tomographic Locality}.
\end{itemize}
This postulate is explained in Subsection~\ref{SubsecTwoGbits}. It is assumed in conjunction with the background assumptions of the GPT framework mentioned there (existence of local preparations and local measurements, and the no-signalling principle).
 \begin{itemize}
\item[3.] \textbf{Existence of an Information Unit}. There exists a fundamental information unit such that the state of any system can be reversibly encoded in a sufficiently large number of these information units, as explained above. It must satisfy these properties\footnote{We skip one of the properties, ``all effects are observable'', since this postulate only applies in a slightly more general situation (where some mathematically possible measurements in the GPT framework may be declared physically forbidden) that we have not introduced in this paper for reasons of brevity.}:
	\begin{itemize}
	\item State tomography is possible: the state space of the information unit is finite-dimensional.
	\item Units can interact: On the state space of \emph{two} information units $A$ and $B$, there exists at least one reversible transformation that does not act independently on $A$ and $B$ (equivalently, at least one locally prepared state $\omega_A\omega_B$ is mapped to a state that cannot be locally prepared).
	\end{itemize}
\end{itemize}
If information units could not interact, then no useful computation would ever be possible: initially uncorrelated universal bits in a state $\omega_A\omega_B$ (in quantum theory, we could write the state as $\rho_A\otimes\rho_B$) would always be transformed to an uncorrelated state $\omega'_A\omega'_B$ (resp.\ $\rho'_A\otimes\rho'_B$, corresponding to a unitary transformation $U_A\otimes U_B$). Not even a classical ``controlled-NOT gate'', for example, could ever be implemented.
\begin{itemize}
\item[4.] \textbf{No Simultaneous Encoding}. If a universal bit is used to perfectly encode one classical bit of information, it cannot be used to simultaneously encode any further information.
\end{itemize}
This postulate says that the name ``universal bit'' is justified, in the sense that the information unit can carry exactly one binary alternative and not more. However, its detailed meaning is somewhat subtle, since interesting non-classical phenomena can happen in the GPT framework. This is most easily expressed by means of an example, namely the gbit from Subsection~\ref{SubsecGbit} (for more details see~\cite{PNAS}). Recall that a gbit state is given by $\omega=(p(\uparrow|0),p(\uparrow|1))$. Now we can encode one bit $b\in\{0,1\}$ by setting $p(\uparrow|0)=b$. This bit can easily be read out: by performing the $0$-measurement on the state $\omega$, we get outcome $\uparrow$ with certainty if $b=1$, and $\downarrow$ with certainty if $b=0$. But we have some more remaining freedom: we can encode an \emph{additional bit} $b'\in\{0,1\}$ into the state, by setting $p(\uparrow|1)=b'$. The bits $b$ and $b'$ determine $\omega$ uniquely; $b'$ can be read out reliably by performing the $1$-measurement.

However, the catch is that we can read out \emph{either} $b$ \emph{or} $b'$, but \emph{not both}: a simple GPT calculation shows that there exists no measurement that reads out both bits reliably at the same time. Moreover, reading out one of the bits introduces unavoidable disturbance that erases the other bit. We have ``complementarity'' in the sense of a ``random access code'': we can decide which \emph{one} of the two bits we would like to read out, but we can never obtain both; hence the name ``gbit''. Yet we can definitely \emph{encode} more than one bit into the state. It is this additional encoding capacity that Postulate 4 intends to forbid --- the gbit violates ``No Simultaneous Encoding''. In addition, it also violates Continuous Reversibility.

Contrasting the quantum to the boxworld case (as done further in Subsection~\ref{SubsecTwoGbits}) hence suggests that the interesting, \textit{uniquely} quantum content of the postulates seems to be contained chiefly in Continuous Reversibility and the interaction of information units. As mentioned in \cite{AlSafi}, it has been conjectured more generally that tomographic locality and (not necessarily continuous!) reversibility are sufficient to pick out only theories embeddable within quantum theory. This would suggest that quantum theory is characterized as the maximal theory with reversible dynamics and tomographic locality. Even in the absence of a proof of this conjecture, it is clear that reversibility plays a key role in quantum theory, and that continuity is what distinguishes quantum theory from classical probability theory over a finite number of (qu)bits.\footnote{This is a remarkable inversion of the standard view that infinite-dimensional \emph{continuous} classical systems can exhibit \emph{discreteness} after quantization.}

Some further remarks regarding the significance of dynamics in singling out quantum theory are necessary. The key roles that continuous reversibility and interaction of information units play highlights the importance of the interplay between dynamics and kinematics \cite{Spekkens2} in physical theory. In order to arrive at the correct state space for a qubit (kinematics), we have to consider the possible interactions (dynamics) that are allowed. This emphasizes Spekkens' point~\cite{Spekkens2} that kinematics and dynamics only play a meaningful physical role when appropriately combined. Individually, each alone is inadequate for a real physical characterization. For example, analysis of static correlation tables from possible quantum experiments \cite{BananaWorld} leads to postulates that allow for \textit{almost-quantum correlations}~\cite{AlmostQuantum} that are slightly more general than those allowed by quantum theory. Perhaps this is a hint at the presence of new physics in regimes where the concept of dynamics becomes inapplicable, such as in quantum gravity, where time may not always be well-defined. In any case, the structure of quantum theory---as we currently know it---depends crucially on the form of permissible dynamics.

We contrast our discussion with that of Fuchs, who anticipated that information-theoretic axiomatizations of quantum theory would show that much---\textit{but not all}---of quantum theory was simply a consequence of information theory~\cite{Fuchs02}. He hoped that what was left would be the true physical kernel of quantum theory, an ambition also expressed by Brukner~\cite[p.\ 23]{Brukner}. Much of the discussion in \cite{Fuchs02} anticipates that contextuality is a genuine physical feature of quantum theory, and will turn out to be the non-information-theoretic kernel. After much progress in axiomatizing quantum theory, Fuchs showed his disappointment with the lack of surprising physical content revealed \cite{Fuchs14}. With a correct axiomatization, he hoped that \emph{``[t]he distillate that remains---the piece of quantum theory with
no information theoretic significance---will be our first unadorned glimpse of `quantum reality.' Far from being the end of the journey, placing this conception of nature in open view will be the start of a great adventure''}~\cite[p.\ 990]{Fuchs02}.

We suggest that (continuous) reversibility may be the postulate which comes closest to being a candidate for a glimpse on the genuinely physical kernel of ``quantum reality''. Even though Fuchs may want to set a higher threshold for a ``glimpse of quantum reality'', this postulate is quite surprising from the point of view of classical physics: when we have a \emph{discrete} system that can be in a \emph{finite} number of perfectly distinguishable alternatives, then one would classically expect that reversible evolution must be discrete too. For example, a single bit can only ever be flipped, which is a discrete indivisible operation. Not so in quantum theory: the state $|0\rangle$ of a qubit can be continuously-reversibly ``moved over'' to the state $|1\rangle$. For people without knowledge of quantum theory (but of classical information theory), this may appear as surprising or ``paradoxical'' as Einstein's light postulate sounds to people without knowledge of relativity.

This viewpoint notwithstanding, our arguments from Section~\ref{SecThesis}, in combination with the result above support the hypothesis that \emph{quantum theory is a principle theory of information}, with continuously-reversible evolution in time as a characteristic property. Any further insights into an underlying ``quantum reality'' (if it exists), or into the question ``information about what'' (if it has an answer) should not be expected to arise directly from these principles, or from quantum theory itself, but from a novel, yet-to-be-found constructive theory with additional beyond-quantum predictive power (if it exists). This is comparable to thermodynamics, whose formulation in terms of principles does not in itself imply the existence of atoms, in contrast to statistical physics which in addition yields novel predictions.

\section{Conclusions}
We take it that information-theoretic reconstructions of quantum theory provide a fruitful---albeit only a \textit{partial}---interpretation of quantum theory. By reconstruing the formalism of quantum theory in terms of operational constraints, one can cast quantum theory as a \textit{principle theory}, and thereby gain explanatory power regarding structural features of a quantum world. In particular, the postulates of most of the current reconstructions (in particular of the one that we have outlined in Section~\ref{SecDerivingQuantumTheory}) are broadly computational and information-theoretic in nature, emphasizing the role that this terminology plays in the formalism of quantum theory. Continuous reversible transformations, which can either be interpreted as computational processes or physical time evolutions, play a major role in singling out quantum theory from the space of GPTs. This indicates that interaction and reversibility of dynamics may be physically characteristic of quantum theory, providing a partial response to the hope that there would be bare physical content separable from the information-theoretic content of quantum theory. However, it is clear that these postulates do not provide a full interpretation; we learn what combination of operational constraints make necessary the structure of quantum theory, but we do not gain a constructive account of ontological structure. Information-theoretic reconstructions can be augmented with further metaphysical claims to arrive at a full interpretation; we outlined three possibilities in Section 2. We also argued that the information-theoretic reconstructions pose a challenge to existing $\psi$-ontic interpretations, by highlighting a lack of explanatory power: these interpretations neither provide a similarly illuminating derivation of the formalism from principles, nor do they yield additional predictions or unification in the empirically accessible realm of our world as one might hope to obtain from a constructive theory.

\section*{Acknowledgments}
We are grateful to Rob Spekkens for helpful discussions. We acknowledge funding from the Canada Research Chairs Program, from the Natural Sciences and Engineering Research Council of Canada (NSERC) Discovery Grants program, and from the Social Sciences and Humanities Research Council of Canada (SSHRC) Joseph-Armand Bombardier Canada Graduate Scholarships. This research was supported in part by Perimeter Institute for Theoretical Physics. Research at Perimeter Institute is supported by the Government of Canada through the Department of Innovation, Science and Economic Development Canada and by the Province of Ontario through the Ministry of Research, Innovation and Science.

\bibliographystyle{alpha}
\bibliography{Bib.bib}

\end{document}